\documentclass[12pt]{iopart}

\usepackage[english]{babel}
\usepackage{iopams}
\usepackage{dsfont}

\expandafter\let\csname equation*\endcsname\relax
\expandafter\let\csname endequation*\endcsname\relax
\usepackage{amsmath}
\usepackage{graphicx}

\usepackage[superscript]{cite}

\begin{document}

\title[Two-dimensional spectroscopy beyond the perturbative limit]{Two-dimensional spectroscopy beyond the perturbative limit: the influence of finite pulses and detection modes}
\author{Andr\'e Anda, Jared H. Cole}
\address{ARC Centre of Excellence in Exciton Science, Australia}
\address{Chemical and Quantum Physics, School of Science, RMIT University, Melbourne, Australia}

\begin{abstract}
Ultra-fast and multi-dimensional spectroscopy gives a powerful looking glass into the dynamics of molecular systems. In particular two-dimensional electronic spectroscopy (2DES) provides a probe of coherence and the flow of energy within quantum systems which is not possible with more conventional techniques. While heterodyne-detected (HD) 2DES is increasingly common, more recently fluorescence-detected (FD) 2DES offers new opportunities, including single-molecule experiments. However in both techniques it can be difficult to unambiguously identify the pathways which dominate the signal. Therefore the use of numerically modelling of 2DES is vitally important, which in turn requires approximating the pulsing scheme to some degree. Here we employ non-pertubative time evolution to investigate the effects of finite pulse width and amplitude on 2DES signals. In doing so we identify key differences in the response of HD and FD detection schemes, as well as the regions of parameter space where the signal is obscured by unwanted artefacts in either technique. Mapping out parameter space in this way provides a guide to choosing experimental conditions and also shows in which limits the usual theoretical approximations work well and which limits more sophisticated approaches are required.
\end{abstract}

\maketitle


\section{Introduction}

The success of 2D spectroscopy derives from its ability to correlate electronic transitions at ultrafast time delays, by achieving high resolutions both in frequency and in time.\cite{10.1088/978-0-750-31062-8,doi:10.1146/annurev.physchem.54.011002.103907,doi:10.1002/andp.201300153,GELZINIS2019271,mukamel1995principles} By portraying the data in a 2D map, coupling between states may be revealed and dark states can be exposed via their coupling to bright states. Moreover, the two dimensions separate homogeneous broadening and inhomogeneous broadening into perpendicular directions, thereby untangling fast fluctuations from the static or slowly fluctuating energy levels.\cite{doi:10.1063/1.3613679} By investigating the time evolution of the spectra, information about the dynamics of the system, including vibrations, relaxation and dephasing, can be obtained.

In multi-pulse spectroscopy, the possibilities explode as the complexity of the experiment increases. However, some designs are more broadly applicable and meaningful and thus establish themselves as routine measurements in labs around the world. As Pump-Probe became the go-to experiment in two-pulse spectroscopy, 2D spectroscopy is becoming the standard measurement with three pulses. 

Pump-Probe, which is the simplest spectroscopic technique used to study ultrafast electronic dynamics, measures the frequency spectrum for a range of pump-to-probe times, and thus provides information on the transient absorption. The extra pulse in 2D spectroscopy enables two frequency spectra to be resolved, one for excitation and one for detection, which are plotted against each other for a range of pulse delays in order to time-resolve the correlation between excitation and detection frequencies.

2D spectroscopy aims to measure the third-order response of a quantum system to an electric field. By clever design of pulses, this goal can be achieved with a high fidelity. First and foremost, the width of the pulse should be short to enhance the time resolution. Equally important is the phase of the pulse which is precisely controlled in order to separate the third-order signal from other orders. 

Since its inception, 2D spectroscopy has employed heterodyne-detection (HD) to record the amplitude and the phase of the signal.\cite{10.1088/978-0-750-31062-8,doi:10.1146/annurev.physchem.54.011002.103907,doi:10.1002/andp.201300153,GELZINIS2019271,hamm_zanni_2011} In HD, the signal is measured by interference with a much stronger electric field, a local oscillator. As HD 2D spectroscopy (HD2D) has matured, so has the understanding of the relationship between the underlying dynamics of the sample and their spectral features. The technique is most widely known for shedding new light on quantum coherent transport in photosynthesis, particularly the hotly debated oscillations in 2D spectra and their origin.\cite{Engel2007,doi:10.1063/1.4846275,Duan_2015,Duan8493} However, HD2D has been shown to be a broadly applicable method.

In recent years, an alternative approach to measure the third-order signal has gained a lot of interest, namely fluorescence-detected 2D spectroscopy (FD2D). Until recently, FD2D was primarily a proof-of-concept method, focusing on experimental developments,\cite{doi:10.1063/1.2800560,doi:10.1063/1.4874697,Draeger:17,Goetz:18}  but it is now establishing itself as an incisive tool with real applications.\cite{Tiwari:18,Tiwari2018,doi:10.1063/1.5046645,C9SC01888C} In the following we focus on comparing HD2D and FD2D; for a broader perspective on multidimensional ultrafast spectroscopy, references \citenum{hamm_zanni_2011}, \citenum{doi:10.1098/rsos.171425} and \citenum{SONG2018184} are excellent resources. 

In this work, we compare the two detection schemes to better understand what they have in common and where they differ. In order to not limit the study to idealised experiments, we chose to go beyond the double-sided Feynman diagrams\cite{mukamel1995principles,doi:10.1063/1.4973975} and simulate the spectra non-perturbatively.\cite{BRUGGEMANN2007192,Br_ggemann_2011} This approach enables more aspects of the experiments to be covered,\cite{Richtereaar7697,Smallwood:17,Perlik:17,doi:10.1063/1.4985888} such as the pulse duration and the pulse amplitude, which are another key focus of our study. However, other sources of divergence bebtween the detection methods could be interesting to investigate, e.g. dephasing and relaxation mechanisms, transition strengths and quantum yields, the effect of vibrations, and the energy level structure. 

\section{Background}

A great advantage of HD2D is that the third-order signal is spatially separated into three spots depending on the type of interaction that occurred between the sample and the electric pulse, with similar phase evolutions interfering constructively in these directions. For historical reasons, these contributios are named the rephasing (R), the nonrephasing (NR) and the double-quantum coherence (DQC) signals.

In FD2D, a fourth pulse, similar to the first three, replaces the local oscillator in the conventional experiment, with the effect that the desired third-order signal is encoded into the excited state populations, from which fluorescence is collected during an acquisition time. The modulation in the integrated fluorescence, as a function of the pulse intervals, are then Fourier transformed to give the 2D spectra which, as for HD2D, can be separated into the R, NR and DQC contributions.

Other actions, such as photoelectron or -ion emission\cite{PhysRevA.92.053412,C5CP03868E} or photocurrent,\cite{Karki2014} could also be recorded to produce 2D spectra with complementary information, but we devote our attention to FD2D as it is currently more popular and also because it affords a more direct comparison to HD2D. 

Detecting the fluorescence instead of the polarisation has a number of potential benefits. The most known advantage is that signals from small volumes, in principle single molecules,\cite{Liebel2018,Tiwari:18} can be used to generate 2D spectra, whereas the conventional technique requires sample sizes larger than the wavelength of the pulse. This opens up the possibility to pierce through the ensemble and study isolated systems or variations across a sample.\cite{Tiwari2018}

In addition, inherent differences between the two detection schemes affect the selection of interaction pathways, which in turn give rise to discrepancies in the spectra. By contrasting HD and FD 2D spectra, it is possible to infer new information which would otherwise be inconclusive with either detection method. Karki et al. compared the HD and FD DQC spectra of LH2 and were thus able to deduce that the initial excitation is shared between the two bacteriochlorophyll rings, contrary to the generally accepted picture up until that point.\cite{C9SC01888C}

Recently, Maly and Mancal suggested that the acquisition time, which has no analogue in HD2D, could be varied in order to isolate specific contributions to the total signal, e.g. the exciton-exciton annihilation.\cite{Maly2018} This opens up a new window into the underlying dynamics, but long measurement times increase the incoherent mixing of linear signals arising from nonlinear population dynamics.\cite{doi:10.1063/1.4994987} However, careful analysis can differentiate between true nonlinear signals and incoherently mixed linear signals.\cite{doi:10.1021/acs.jpca.9b01129}

Experimentally, FD2D is currently more demanding than its counterpart. This is partly because two pulse delays need to be scanned instead of one. Moreover, labs which have implemented the fluorescence-detection setup are relatively few, and don't enjoy the same level of experience and literature to draw upon. However, the increasing rate of articles published suggest that FD2D is leaving infancy and is becoming a powerful addition to the toolbox.

Only a handful of papers have tackled the theory side of FD2D.\cite{Perdomo-Ortiz2012,doi:10.1021/acs.jpca.9b01129,PhysRevA.96.053830,Maly2018,khn2019interpreting} Using the double-sided Feynman diagrams derived from perturbation theory, it is readily shown that each diagram from the traditional method has an equivalent diagram in fluorescence detection, which in addition has an extra set of excited state absorption (ESA) diagrams.\cite{Perdomo-Ortiz2012} Although this knowledge forms an important basis for the interpretation of FD2D spectra, the differences do not end there.

Knowing how challenging it can be to analyse traditional 2D spectra, where the various contributions to the signal are well understood, it is of great interest to explore all possible ways that FD2D spectra can deviate from its counterpart, in order to fully unlock the potential of the method.

In order to enhance the comparison between HD and FD, and to clarify the interpretation of the 2D spectra, we make some simplifications. Whereas the FD2D experiment requires two pulse intervals to be scanned, and the HD2D version only one, we chose to scan both in our simulations. Also, the coherent detection in the HD experiment is performed by taking the instantaneous expectation value of the transition dipole moment, we do not model the local oscillator with a nonequilibrium Green's function QED approach.\cite{PhysRevA.77.022110} Both of the abovementioned choices are in line with how the perturbative simulations using double-sided Feynman diagrams are typically performed. 

Moreover, as FD2D only requires a single absorber to generate spectra, we do not include a distribution of energy levels as a source of inhomogeneous broadening. For the same reason, we disregard the vector property of the transition dipole moment and treat them as scalars. To promote as identical conditions as possible, we do the same for the HD2D simulations, which \emph{do} require an ensemble of absorbers - however, only the positions are different. 

In our simulations, we use the rotating frame instead of the laboratory frame or the quasi-rotating frame.\cite{Kramer:16} We also sample uniformly, opting for simplicity rather than trying to reduce the computational cost with non-uniform sampling.\cite{doi:10.1063/1.4976309,wang2019compressed} We do not investigate the polarisation dependency, which is sometimes exploited to select specific pathways\cite{Zhang14227,Stone1169,Mueller2018,Thyrhaug2018,Kramer2020}, but it is known that the early-time dynamics suffer from incorrect pulse-ordering artifacts.\cite{doi:10.1063/1.5079817}

We assume non-interacting chromophores in our calculations, although the delocalisation of excitation energy in many cases is unexpectedly long-range\cite{C8CP05851B} and its role in energy transfer is a hot topic in the field.\cite{Strumpfer2012,doi:10.1063/1.5046645,C5CP06491K}




\section{Theory}

First we introduce the theoretical and computational methods used throughout the article, starting with the evolution of a quantum system interacting semiclassically with an electric field. We then describe the equations used to detect the third-order signal and construct the 2D spectra for both heterodyne and fluorescence detection. Lastly, we explain the model system which we use in our simulations.

\subsection{The quantum dynamics of a system interacting with an electric field} 

The vast majority of 2D spectroscopy models employ the semiclassical approximation, where the state of the system is described quantum mechanically, but the field, $E(\mathbf{r},t)$, is described classically:

\begin{equation}
E(\mathbf{r},t) = \sum_n A_n(t-t_n) \mathrm{exp}(-i\omega (t-t_n) + i\mathbf{k}_n\cdot \mathbf{r} - i\phi_n) 
\end{equation}

\noindent Here, $A_n(t-t_n)$ describes the envelope of the $n$th pulse centred at $t_n$; $\omega$ is the frequency of the field; $\mathbf{k}_n$ is the wave vector of the $n$th pulse; and $\phi_n$ is the phase angle of the $n$th pulse. 

The coupling between the quantum state and the classical field is given by the interaction Hamiltonian

\begin{equation}
H_{\mathrm{int}}(t) = -\mu E(t)
\end{equation}

\noindent where $\mu$ is the transition dipole moment operator, which is assumed to be a scalar for simplicity. For an otherwise isolated quantum system, the dynamics obeys the time-dependent Schrodinger equation

\begin{equation} i\hbar \frac{\partial}{\partial t} | \Psi (\mathbf{r}, t) \rangle = [ H_0 + H_{int}(t)]  | \Psi (\mathbf{r}, t) \rangle \equiv H(t)  | \Psi (\mathbf{r}, t) \rangle 
\end{equation}

\noindent where $H_0$ is the system Hamiltonian in the absence of the field. In the condensed phase, however, the fluctuations in the quantum system's environment perturb the system sufficiently that an isolated-quantum system theory is inadequate to describe the dynamics, as it can not include relaxation and dephasing processes. To accommodate these effects, an open-quantum system approach is needed. For the sake of simplicity, we use the Lindblad equation to propagate the state, which is now represented by the reduced density operator, $\rho(t)$.\cite{Lindblad1976}

\begin{align}\label{eq:generalLindblad}
\frac{d}{dt}\rho(t) = &\frac{i}{\hbar}[\rho(t),H(t)] + \sum_{j=1}^{N}\Gamma_j \big \{L_j\rho(t)L_j^\dagger - \frac{1}{2}[\rho(t)L_j^\dagger L_j + L_j^\dagger L_j \rho(t)]\big \} \\ \equiv & \mathcal{L}[\rho (t)] \nonumber
\end{align} 

\noindent where an off-diagonal $L_j = a_m^\dagger a_n$ operator represents spontaneous relaxation or excitation,  a diagonal $L_j$ operator represents a dephasing process, and $N$ denote the number of different decoherence channels. The formal solution to the differential equation in (\ref{eq:generalLindblad}) is found by integrating both sides of the equation: $\rho(t) = e^{\mathcal{L}t}\rho(0)$.


Having established the dynamical equations for an open quantum system interacting with a classical field, we proceed to discuss the different detection schemes for the third-order signal. Traditionally, heterodyne-detection has been used, where the emitted light is interfered with a local oscillator and the electric field is measured. More recently, fluorescence-detection has become an increasingly viable and attractive method, as technological advances push towards single-molecule 2DES experiments, exploiting the high sensitivity of fluorescence detectors.

\subsection{Heterodyne-Detected 2DES}

Heterodyne-Detected 2DES employs phase matching to isolate the third-order response signal. In short, three pulses impart their unique wave vectors to the system, thus causing the radiated fields from individual emitters to constructively interfere in the direction of the vector sum $\mathbf{k}_{signal} = \mp \mathbf{k}_{1} \pm \mathbf{k}_{2} + \mathbf{k}_{3}$, where the upper combination corresponds to the rephasing signal and the lower combination to the nonrephasing signal. A third (possible) phase matching direction, $\mathbf{k}_{\mathrm{DQC}} =  \mathbf{k}_{1} + \mathbf{k}_{2} - \mathbf{k}_{3}$, produces the so-called double-quantum coherence signal, but this will not be considered in our study.

Importantly, a detectable signal in the phase-matched directions relies on an ensemble of oscillating dipoles to coherently add up to a macroscopic polarisation in the sample. While this enables high signal-to-noise ratios, it also puts a lower bound on the spatial resolution: $\gtrsim \lambda ^3$, where $\lambda$ is the radiation wavelength.


Each optically active molecular system, $j$, picks up a unique phase factor by virtue of its position, $\mathbf{r}_j$, since the electric field is given by

\begin{equation} E_j(t) = E_0\sum_{n=1}^3 \mathrm{exp}\bigg [\frac{t-t_n}{2\sigma^2}\bigg ]\mathrm{exp}(i\omega t - i \mathbf{k}_n\cdot \mathbf{r}_j)
\end{equation}

\noindent where a Gaussian envelope function is used and the $\phi_n$ phase has been dropped as it is not necessary for phase matching. The electric field couples to the transition dipole moment of the system which makes the $\mathrm{exp}(- i \mathbf{k}_p\cdot \mathbf{r}_j)$ phase factor act as a book-keeper of the transitions. By using linearly independent $\mathbf{k}$-vectors for each of the three pulses, it is possible to differentiate which transitions were caused by which pulses.

\begin{figure}
\includegraphics[width=0.5\textwidth]{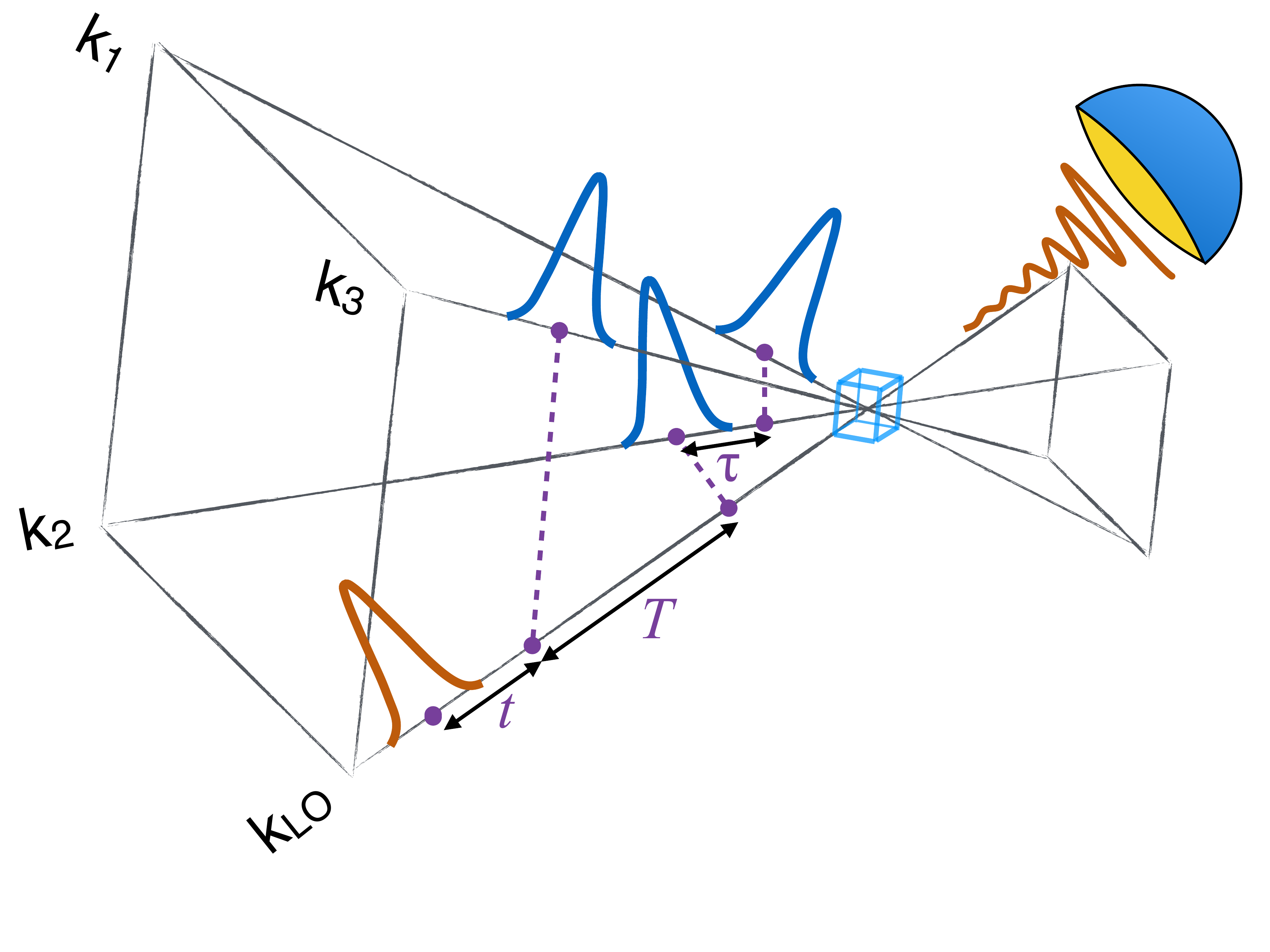}
\put(-230,155){(a)}
\includegraphics[width=0.5\textwidth]{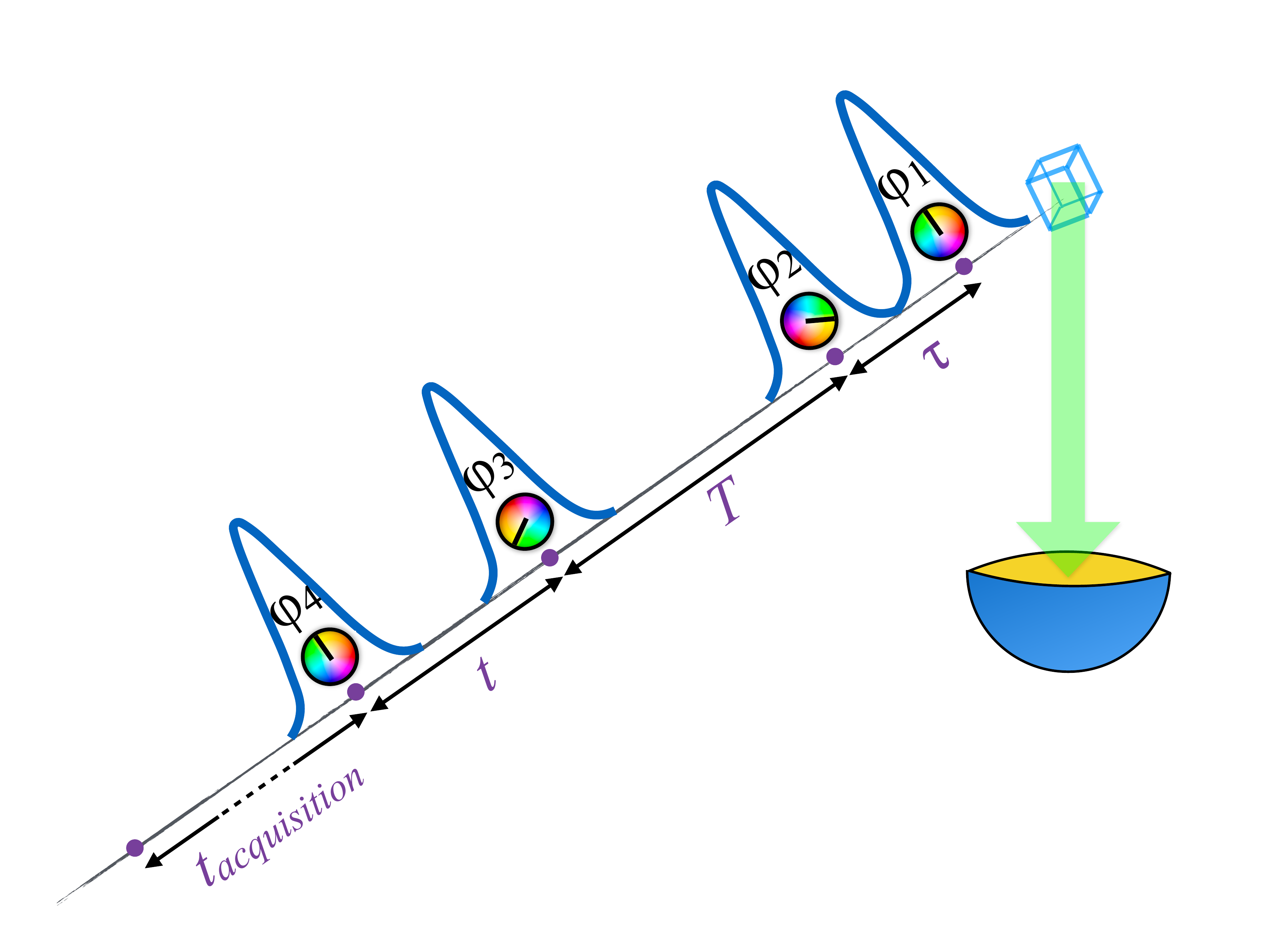}
\put(-210,155){(b)}
\caption{\label{fig:HDFDsetup}(a) Heterodyne-detection of two-dimensional electronic spectra employs a local oscillator (LO) to read out the macroscopic polarisation generated by three interactions with the electric field. In the non-collinear geometry, linearly independent wave vectors imprint unique phases with each photo-excitation and de-excitation. The selection of specific third-order processes, such as rephasing and nonrephasing, is achieved through phase-matching, in which the wave vector of the LO is a combination of the previous wave vectors, but with the necessary signs flipped to reflect the order of (de-)excitation for the respective processes. (b) Fluorescence-detection of two-dimensional electronic spectra is performed in a collinear setup. Instead of measuring the polarisation, fluorescence is detected up to a cut-off acquisition-time. The integrated fluorescence as a function of the pulse delays constitutes the signal. However, the selection of third-order processes requires each pulse train to be executed 27 times with phases cycled. Subsequent Fourier transforms with respect to the distinct phase evolutions of rephasing and nonrephasing processes, produce the 2D spectra which are related to, but not equivalent to, the heterodyne-detected 2D spectra.}
\end{figure}

However, averaging over the polarisation of the individual molecules will cause the third-order signal to vanish due to the random phases, unless the polarisation is measured in one of the phase-matching directions. That is, to read out the desired signal, the wave vector of the local oscillator, $\mathbf{k}_{signal}$, must be chosen such that it simultaneously cancels the phase factors absorbed by all molecular systems for the relevant pathways. For 2DES, we are interested in the part of the response which stems from a single interaction with each of the three pulses. This reduces the number of phase-matching directions to three: $- \mathbf{k}_{1} + \mathbf{k}_{2} + \mathbf{k}_{3}$, $\mathbf{k}_{1} - \mathbf{k}_{2} + \mathbf{k}_{3}$ and $\mathbf{k}_{1} + \mathbf{k}_{2} - \mathbf{k}_{3}$. A schematic of the phase matching method is shown in figure \ref{fig:HDFDsetup}(a).


To achieve the phase-matching in a non-perturbative calculation, the dynamics of each individual molecule $j$ must be computed separately using equation~\ref{eq:generalLindblad}. Here, we assume that the individual molecules evolve independently and that each molecule is initially in its ground state. For simplicity, we employ identical Hamiltonian and Lindblad operators on each individual quantum system. A generalisation to a distribution of Hamiltonians and Lindbladians is straightforward and with no added computational cost, however, this would be a source of both inhomogeneous and homogeneous broadening which could potentially make it more difficult to interpret (the differences between) the HD and FD 2D spectra. 

The polarisation in the chosen phase-matching direction is given by:

\begin{equation} \label{eq:detectPolarisation}
P(\mathbf{k}_{signal},\tau, T, t) = 2\mathrm{Re}\sum_j \mathrm{exp}(i \mathbf{k}_{signal} \cdot \mathbf{r}_j)\sum_{\alpha < \beta} \mu_{\alpha \beta}^{(j)} \rho_{\beta \alpha}^{(j)} (E_j;t)
\end{equation}  

\noindent The last step is to perform a two-dimensional Fourier transform from the time domain to the frequency domain:

\begin{equation} S_{\mathrm{HD2D}}(\mathbf{k}_{signal},\omega_\tau, T, \omega_t) = -i \int_{0}^{\infty} \mathrm{d}\tau \mathrm{exp}(\pm i \omega_\tau \tau ) \int_{0}^{\infty} \mathrm{d}t \mathrm{exp}(i \omega_t t) P(\mathbf{k}_{signal},\tau, T, t) 
\end{equation}

\noindent where ``+'' is used for the nonrephasing and double-quantum coherence signals and ``-'' is used for the rephasing signal. In the following, only the real parts of the signals will be investigated, as this information is far more often used in studies - the imaginary part is mostly neglected.

\subsection{Fluorescence-Detected 2DES}

Instead of detecting the third-order polarisation with a local oscillator, the integrated fluorescence intensity can be used to report on the nonlinear response of the sample. 
Fluorescence-detection is extremely sensitive, which allows for studies of single molecules. As fluorescence is an incoherent process, no phase-matching condition can exist, and there is therefore no need to perform the experiment in a noncollinear setup, which is standard for heterodyne-detected 2DES. The advantage of the collinear geometry is that it facilitates rapid data acquisition and that it is inherently phase stable. Two major drawbacks are the need to scan two coherence times and that it is necessary to scan a number of different pulse phases for each unique combination of pulse delays,  in order to extract the desired rephasing and nonrephasing signals. The latter requirement is achieved by two similar but different methods known as phase modulation\cite{doi:10.1063/1.2800560}, which will not discussed here, and phase cycling.\cite{Tian1553} 






\subsubsection{Phase Cycling}

In the collinear setup, the signal is no longer spatially separated according to the order of the light-matter interaction; it now contains all orders. To isolate the rephasing and nonrephasing (and DQC) signals, one must exploit the unique phase evolutions of the third-order signals. This can be done by Fourier transforming the signal with respect to the phase for each pulse interval, so that it matches the characteristic phases of the R, NR and DQC signals. Practical limitations prevent a continuous Fourier transform, but it can be shown that 27 different phase combinations is enough to extract the third-order signals.\cite{doi:10.1063/1.2978381} Figure~\ref{fig:HDFDsetup}(b) is a schematic of the FD2D experiment emphasising the collinear geometry and the phase-cycling approach, where the pulses are tagged with specific phases which are then rotated or cycled with respect to each other.

\subsubsection{Example with excited state population}

From a purely theoretical point of view, the excited state populations have the same dependence on the interactions with the four pulses as the integrated fluorescence, meaning that the populations can be used as proxies for the measured signal.\cite{doi:10.1021/acs.jpca.9b01129} In principle, one can distinguish between different excited state populations, but since it is difficult to detect fluorescence from distinct excited states, we neglect this possibility.\cite{PhysRevA.96.053830} It is, however, relevant for other action spectroscopies.\cite{Karki2014,PhysRevA.92.053412,C5CP03868E}

The excited state population after interactions with four pulses with phases $\phi_{1-4}$ and delay times $\tau$, $T$ and $t$ can be found by summing all the contributing coherence transfer pathways:

\begin{equation} \label{eq:pop1} p(\tau,T,t,\phi_1,\phi_2,\phi_3,\phi_4) = \sum_{\alpha,\beta,\gamma,\delta}\tilde{p}(\tau,T,t,\alpha,\beta,\gamma,\delta)\mathrm{exp}[i(\alpha \phi_1 + \beta \phi_2 + \gamma \phi_3 + \delta \phi_4)] 
\end{equation}
Here, $\alpha,\beta,\gamma,\delta$ count the number of positive phase factors minus the number of negative phase factors imparted on the system by the respective pulses, which in the language of double-sided Feynman diagrams amounts to the number of arrows pointing to the left minus the number of arrows pointing to the right. $\tilde{p}$ denotes that the contributions are specific to the particular set of $\alpha,\beta,\gamma,\delta$ and that the absorbed phase factors have been taken out.

Because it is assumed that the initial state is diagonal, $\rho_0 = |g\rangle \langle g |$, and the final state is diagonal, the following condition must be fulfilled: $\alpha + \beta + \gamma + \delta = 0$. Consequently, $\alpha$, $\beta$, $\gamma$, $\delta$  are dependent, which allows us to define $\phi_{21} \equiv \phi_2 - \phi_1$, $\phi_{31} \equiv \phi_3 - \phi_1$, $\phi_{41} \equiv \phi_4 - \phi_1$. Equation \eqref{eq:pop1} then becomes:

\begin{equation}
p(\tau,T,t,\phi_{21},\phi_{31},\phi_{41}) = \sum_{\beta,\gamma,\delta}\tilde{p}(\tau,T,t,\beta,\gamma,\delta)\mathrm{exp}[i(\beta \phi_{21} + \gamma \phi_{31} + \delta \phi_{41})]  \label{eq:pop2} 
\end{equation}

\noindent However, for the 2DES signals, we are not interested in expressing the total population as a sum of its contributions with unique phase factors. Instead, we wish to isolate the individual contributions, in particular the rephasing, nonrephasing and double coherence signals, which are given by $\tilde{p}(\beta=1, \gamma=1,\delta=-1)$, $\tilde{p}(\beta=-1, \gamma=1,\delta=-1)$ and $\tilde{p}(\beta=1, \gamma=-1,\delta=-1)$, respectively. This is achieved by Fourier transforming the total population with respect to the characteristic phase evolutions of the third-order signals.

\begin{equation}\tilde{p}(\tau,T,t,\beta, \gamma,\delta) = \frac{1}{(2\pi)^3} \int_0^{2\pi} \int_0^{2\pi} \int_0^{2\pi} d\phi_{41} d\phi_{31} d\phi_{21} p(\tau,T,t,\phi_{21},\phi_{31},\phi_{41})e^{-i\beta \phi_{21}}e^{-i\gamma \phi_{31}}e^{-i\delta\phi_{41}}
\end{equation}

\noindent As noted above, a continuous Fourier transform is impractical, so a discrete Fourier transform is used.

\begin{equation}\label{eq:discreteFT}
\tilde{p}(\tau, T, t,\beta, \gamma,\delta) = \frac{1}{LMN}\sum_{n=0}^{N-1}\sum_{m=0}^{M-1}\sum_{l=0}^{L-1}p(\tau,T,t,l\Delta \phi_{21},m\Delta \phi_{31},n\Delta \phi_{41})e^{-il\beta \Delta \phi_{21}}e^{-im\gamma \Delta \phi_{31}}e^{-in\delta \Delta \phi_{41}}
\end{equation}

As shown by Howe-Siang Tan,\cite{doi:10.1063/1.2978381} a phase-cycling scheme of $L\times M\times N = 3\times3\times3$ is sufficient when only contributions from fourth order and below are considered, i.e. $|\alpha| + |\beta| + |\gamma| + |\delta| \leqslant 4$. This gives $\Delta \phi_{21} = \Delta \phi_{31} = \Delta \phi_{41} = \frac{2\pi}{3}$. 

It follows that the appropriate electric field needed to perform the phase-cycling to extract the R, NR, and DQC signals is

\begin{equation} E^{lmn}(t) = E_1(t) + E_2^l(t) + E_3^m(t)  + E_4^n(t) 
\end{equation}

\begin{equation} 
\begin{split}E_1(t) & = E_0 \mathrm{exp}\big [\frac{t-t_1}{2\sigma^2}\big ]^2\mathrm{exp}(i\omega t)\\
E_2^l(t) & = E_0 \mathrm{exp}\big [\frac{t-t_2}{2\sigma^2}\big ]^2\mathrm{exp}(i\omega t + il\tfrac{2\pi}{3}) \\
E_3^m(t) & = E_0 \mathrm{exp}\big [\frac{t-t_3}{2\sigma^2}\big ]^2\mathrm{exp}(i\omega t + im\tfrac{2\pi}{3}) \\
E_4^n(t) & = E_0 \mathrm{exp}\big [\frac{t-t_4}{2\sigma^2}\big ]^2\mathrm{exp}(i\omega t + in\tfrac{2\pi}{3})
\end{split}
\end{equation}

\noindent where $l,m,n$ are cycled between 0,1 and 2 and the resulting populations from the 27 combinations added for each set of \{$t_1,t_2,t_3,t_4$\}. Note that the $\mathbf{k}_n$ vector is dropped as the $\mathbf{k}_n \cdot \mathbf{r}$ phase factors cancel out in the collinear geometry. The R, NR, and DQC spectra are then found using equation \ref{eq:discreteFT} and performing the usual 

\small

\begin{equation} S_{\mathrm{FD2D}}(\beta, \gamma,\delta,\omega_\tau, T, \omega_t) = -i \int_{0}^{\infty} \mathrm{d}\tau \mathrm{exp}(\pm i \omega_\tau \tau ) \int_{0}^{\infty} \mathrm{d}t \mathrm{exp}(i \omega_t t) \tilde{p}(\beta, \gamma,\delta,\tau, T, t)
\end{equation}

\normalsize

\noindent and inserting the appropriate $\beta, \gamma, \delta$ for the R, NR, and DQC signals. The sign of the Fourier transforms are the same as for the HD2D case.

It should be noted, however, that higher-order processes can have identical phase evolutions as the third-order processes; for example, three interactions with the first pulse can leave the system in the same state as only one interaction with the pulse will. The phase-cycling operation will not be able to filter out these higher-order contributions and the spectra will become distorted as a consequence. Care should therefore be taken to ensure that the third-order contributions are predominant, both in experiments and in simulations. 


\subsubsection{Using the fluorescence signal}

Instead of using the final populations as proxies\cite{doi:10.1021/acs.jpca.9b01129} to report on the molecular response, the integrated fluorescence can be recorded. This is also more in line with the actual experiment. Theoretically, the integrated fluorescence can be found as the integral of the spontaneous relaxation which is given by

\begin{equation} \label{eq:detectRelax}
\mathrm{Rel}_{10} = \int_0^{t_{acq}}dt \Gamma_{10} \mathrm{Tr} [L_{10}\rho_S(t)L_{10}^\dagger]
\end{equation} 

\begin{equation}
L_{10} = a_0^\dagger a_1 
\end{equation}

\noindent and similarly for relaxations between other states. $t_{acq}$ denotes the acquisition time in which fluorescence is collected. If it is possible to distinguish particular transitions, which is the case for some action spectroscopies, they can give rise to multiple 2D spectra and provide even more incisive information on the studied system. Otherwise the total integrated fluorescence is just the sum of the individual transitions. Note that the acquisition time can be varied which opens up another window into the dynamics of the sample.

Once the spontaneous relaxation has been recorded, the data undergoes the same phase cycling process as for the population data to construct the rephasing and nonrephasing spectra.

\subsection{The Diagrammatic Approach to 2DES}

The theoretical framework for both the HD and FD spectra can be simplified substantially by invoking a few approximations, resulting in a set of equations each representing unique spectroscopic pathways. These are commonly depicted as double-sided Feynman diagrams (DSFD) which provide a visual connection to the physical processes and are constructed by following a list of rules dictated by the topology of the model system. A neat property of the DSFD equations is that the Fourier transforms can be calculated analytically in Liouville space,\cite{C7CP06583C} 

\begin{align}\label{FTliouville}
 \mathcal{G}^{\pm}(\omega, \tau_f ) \equiv \int_0^{\tau_f} e^{\pm i\omega t}\mathcal{G}(t)dt = [\pm i\omega \mathds{1} + \mathcal{L}]^{-1}(e^{\pm i\omega \tau \mathds{1}+\mathcal{L}\tau} - \mathds{1}) 
\end{align}

\noindent as each propagator $\mathcal{G}(t) \equiv e^{\mathcal{L}t} $ is independent of other time variables. This obviates the cumbersome process of numerically integrating the state of the system for all realisations of the pulse intervals.

\begin{figure}
\centering
         \includegraphics[width=0.8\linewidth]{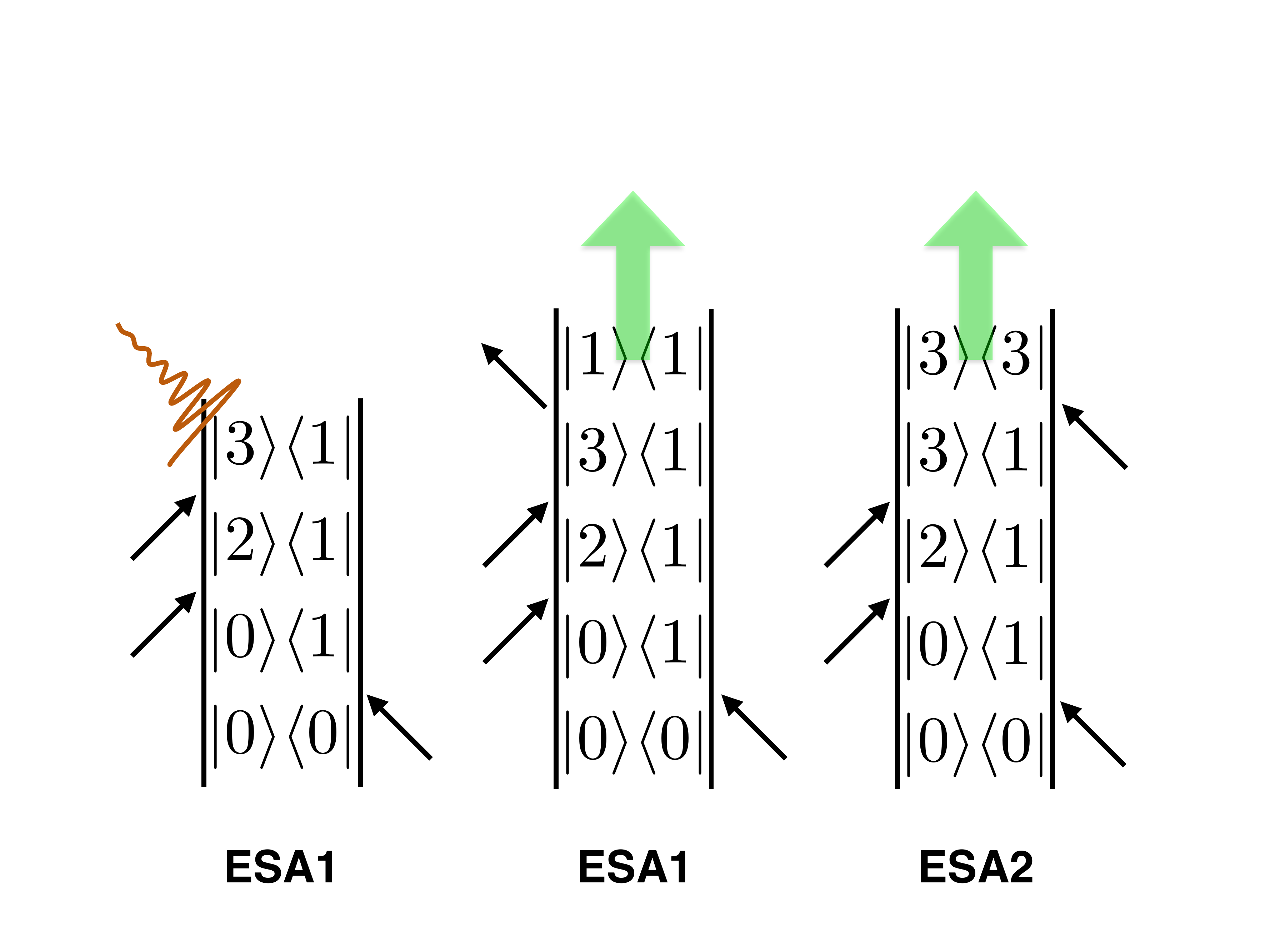}
\caption{\label{fig:esa} Double-sided Feynman diagrams of an identical excited state absorption process, but with different read-outs due to the choice of detection method. Traditional HD (left) interferes a local oscillator with the final coherence to produce a signal. FD, on the other hand, interacts with a fourth pulse which can result in a population of a lower excited state (middle) or a higher excited state (right), from which emission of fluorescence is continually detected. The middle diagram shares the same phase evolution as the HD diagram to the left (both are ESA1) as indeed every HD diagram has an equivalent FD diagram. The rightmost diagram, however, is unique to FD, and is denoted ESA2. Owing to the opposite sign acquired in the last pulse interaction, the middle and the right diagrams largely cancel, which becomes the main source of discrepancy between HD and FD spectra.}
\end{figure}

More detailed accounts on the subject can be found elsewhere, but we include three diagrams which highlight the difference between HD and FD. Figure~\ref{fig:esa} shows excited state absorption diagrams which evolve identically until the detection event as is also evident from their corresponding equations:

\small

\begin{align}
& \mathrm{ESA1_{HD}}(\omega_1,t_2, \omega_3) =  \label{HDdiagram} \mathrm{Tr} \big [ \overrightarrow{\mathcal{V}_{03}}\mathcal{G}^{+}( \omega_3) \overrightarrow{\mathcal{V}_{32}}\mathcal{G}( t_2) \overrightarrow{\mathcal{V}_{20}}\mathcal{G}^{-}( \omega_1) \overleftarrow{\mathcal{V}_{01}}|0\rangle \langle 0 |  \big ]
\\ \nonumber 
\\
& \mathrm{ESA1_{FD}}(\omega_1,t_2, \omega_3;t_{acq}) = \label{FDdiagram1} \int_0^{t_{acq}}dt\gamma_1 \mathrm{Tr}\big [|1\rangle \langle 1 |   \mathcal{G}( t)\overrightarrow{\mathcal{V}_{13}}\mathcal{G}^{+}( \omega_3) \overrightarrow{\mathcal{V}_{32}}\mathcal{G}( t_2) \overrightarrow{\mathcal{V}_{20}}\mathcal{G}^{-}( \omega_1) \overleftarrow{\mathcal{V}_{01}}|0\rangle \langle 0 | \big ] 
\\ \nonumber 
\\
& \mathrm{ESA2_{FD}}(\omega_1,t_2, \omega_3;t_{acq}) = \label{FDdiagram3} \int_0^{t_{acq}}dt\gamma_3 \mathrm{Tr}\big [|3\rangle \langle 3 | \mathcal{G}( t)\overleftarrow{\mathcal{V}_{13}}\mathcal{G}^{+}( \omega_3) \overrightarrow{\mathcal{V}_{32}}\mathcal{G}( t_2) \overrightarrow{\mathcal{V}_{20}}\mathcal{G}^{-}( \omega_1) \overleftarrow{\mathcal{V}_{01}}|0\rangle \langle 0 |\big ] 
\end{align}

\normalsize

\noindent Here, $\overrightarrow{\mathcal{V}_{ij}}$ denotes the transition dipole moment operator acting on the right, causing a transition from $j$ to $i$: $\overrightarrow{\mathcal{V}_{ij}} |j\rangle \langle \bullet | = |i\rangle \langle \bullet |$. Conversely, $\overleftarrow{\mathcal{V}_{ij}}$ acts on the left, causing a transition from $i$ to $j$: $\overleftarrow{\mathcal{V}_{ij}} |\bullet \rangle \langle i | = |\bullet \rangle \langle j |$ The fluorescence yields are given by $\gamma_i$, and $t_{acq}$ is the acquisition time. The initial state, on the right of the equations, is assumed to be the ground state, hence $|0\rangle \langle 0 |$, but can be completely general if desired.

In theory, dropping the integral and taking only the projected populations would give perfectly valid third-order signals, however, this would not be in line with the experiment. Additionally, the acquisition time parameter can be exploited to reveal more information about the system.

Using FD, the last pulse interaction can bring the state to two different excited state populations: The lower excited state, equation~\eqref{FDdiagram1}, which is equivalent to the HD diagram, equation~\eqref{HDdiagram}, and the higher excited state, equation~\eqref{FDdiagram3}, which has no HD counterpart. Moreover, the two FD diagrams have opposite signs resulting in cancellations of these contributions. Depending on the model system, this can lead to peaks appearing with one detection method, but not the other.

For future reference, we note that rephasing (R) designates contributions/diagrams with opposite phase evolutions after the first and third pulse, whereas nonrephasing (NR) contributions/diagrams oscillate with the same sign in these periods. Double-quantum coherence (DQC) contributions/diagrams are in this sense a special case of NR, with the distinction that they are not static or slowly evolving after the second pulse, but oscillate with double frequency.

\subsection{Model} 
\begin{figure}
\centering
         \includegraphics[width=0.7\linewidth]{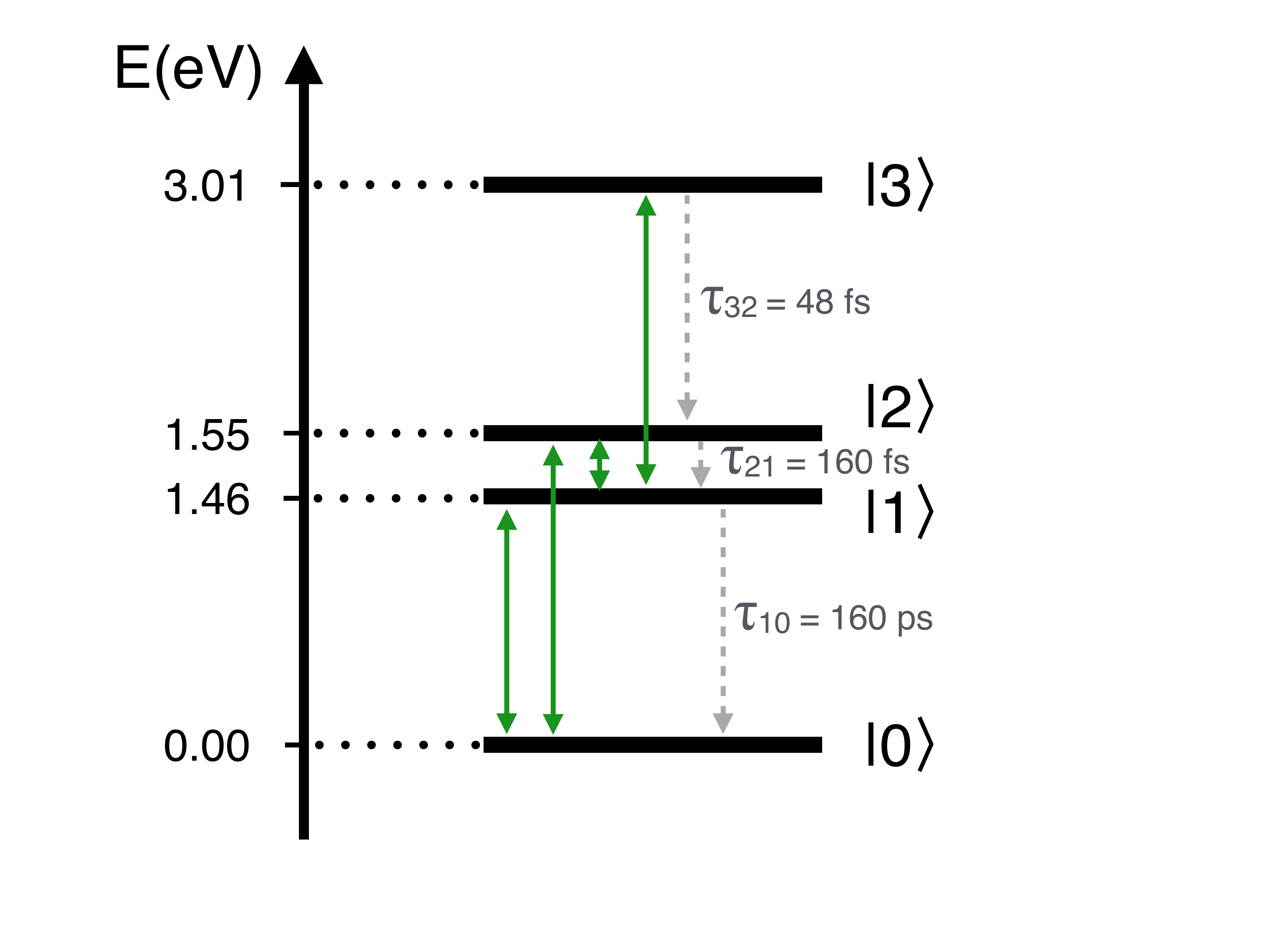}
\caption{\label{fig:model} The model system has four energy levels where all transitions are optically allowed and equally strong, apart from $|0\rangle \rightarrow |3\rangle$, which is forbidden. The relaxation rates $\tau_{ij}$ indicate an ultrafast relaxation pathway from the highest-excited state, and a slightly slower relaxation from the second excited state. The decay from the lowest-excited state to the ground state is orders of magnitude slower. This model system is identical to the one in ref. \citenum{PhysRevA.96.053830}.}
\end{figure}

In figure \ref{fig:model} we depict our model system, which we adopted from reference \citenum{PhysRevA.96.053830}. This four-level system captures a lot of the physics of 2D spectroscopy without complicating the interpretation unnecessarily. It also encompasses the case of an excitonic dimer, which is of fundamental interest and a natural starting point for a comparison between heterodyne-detection and fluorescence-detection. Moreover, the fact that the authors of reference \citenum{PhysRevA.96.053830} simulated the FD2D spectrum using the phase-modulation technique, and not phase-cycling, allows us to validate our calculations and compare the two 
methods of extracting the desired signal. 
We use the same values for the energy levels as given in reference \citenum{PhysRevA.96.053830}:


\begin{align}
&E_0 = 0 \\
&E_1 = 1.46 \: \mathrm{eV}\nonumber \\
&E_2 = 1.55 \: \mathrm{eV} \nonumber \\
&E_3 = E_1 + E_2 \nonumber
\end{align} \label{levels}

\noindent where $E_1$ and $E_2$ correspond to the B850 and B800 absorption bands of the inner and outer rings of bacteriochlorophylls in the light-harvesting complex of purple bacteria.\cite{McDermott1995,Anda2016,Anda2017,DeVicoE9051,Anda2019} The optical transitions between the energy levels are all allowed and are equally strong: $eE_0\mu_{ij} = 8 \: \mathrm{meV} $, with the exception of the transition from ground to the highest excited state, which is forbidden. Electronic relaxation is modelled by the jump operators, the off-diagonal Lindblad operators: $L_{10} = |0\rangle \langle 1 |$, $L_{21} = |1\rangle \langle 2 |$ and $L_{32} = |2\rangle \langle 3 |$, which are scaled by $\Gamma_{10} = 4.13 \: \mu \mathrm{eV}$, $\Gamma_{21} = 4.13 \: \mathrm{m eV}$, $\Gamma_{32} = 13.78 \: \mathrm{m eV}$ respectively. This yields relaxation times ($=\hbar / \Gamma_j$) of 160 ps, 160 fs and 48 fs, which is representative of a dimer system. Dephasing is included similarly with diagonal Lindblad operators: $L_{00} = |0\rangle \langle 0 |$, $L_{11} = |1\rangle \langle 1 |$, $L_{22} = |2\rangle \langle 2 |$ and $L_{33} = |3\rangle \langle 3 |$, which are all scaled by the same dephasing strength $\Gamma_{Dephasing} = 41.3 \: \mathrm{meV}$.

The model system is initialised with the ground state fully populated and then propagated according to equation \ref{eq:generalLindblad} with the appropriate (light-matter) interaction Hamiltonian until it is detected as described by equation \ref{eq:detectPolarisation} or \ref{eq:detectRelax}. As FD only requires an individual photoactive system, in contrast to HD which requires an ensemble, we chose to neglect the distributions of energy levels and transition dipole moments as they potentially could obscure the effects we want to study. However, the computational cost of incorporating these details into our model would be negligible.

In order to ensure a random sampling of the position-dependent phase, it is necessary to sample a volume of space with dimensions greater than the wavelength of the incident field. This is achieved by randomly generating the molecular positions and multiplying with a factor which make intermolecular distances large compared to the wavelength of the pulses.

\section{Results}

By applying the theory of HD and FD 2DES on the model system, we are able to non-perturbatively simulate 2D spectra under a range of different conditions and to study effects that will be applicable to a broad range of samples and experiments. Our goal is to investigate under which limits the extraction of the third-order signal breaks down due to pulse intensity or pulse duration, and hence to understand the optimal regime for 2D spectra.

\begin{figure}
\centering
         \includegraphics[width=0.8\linewidth]{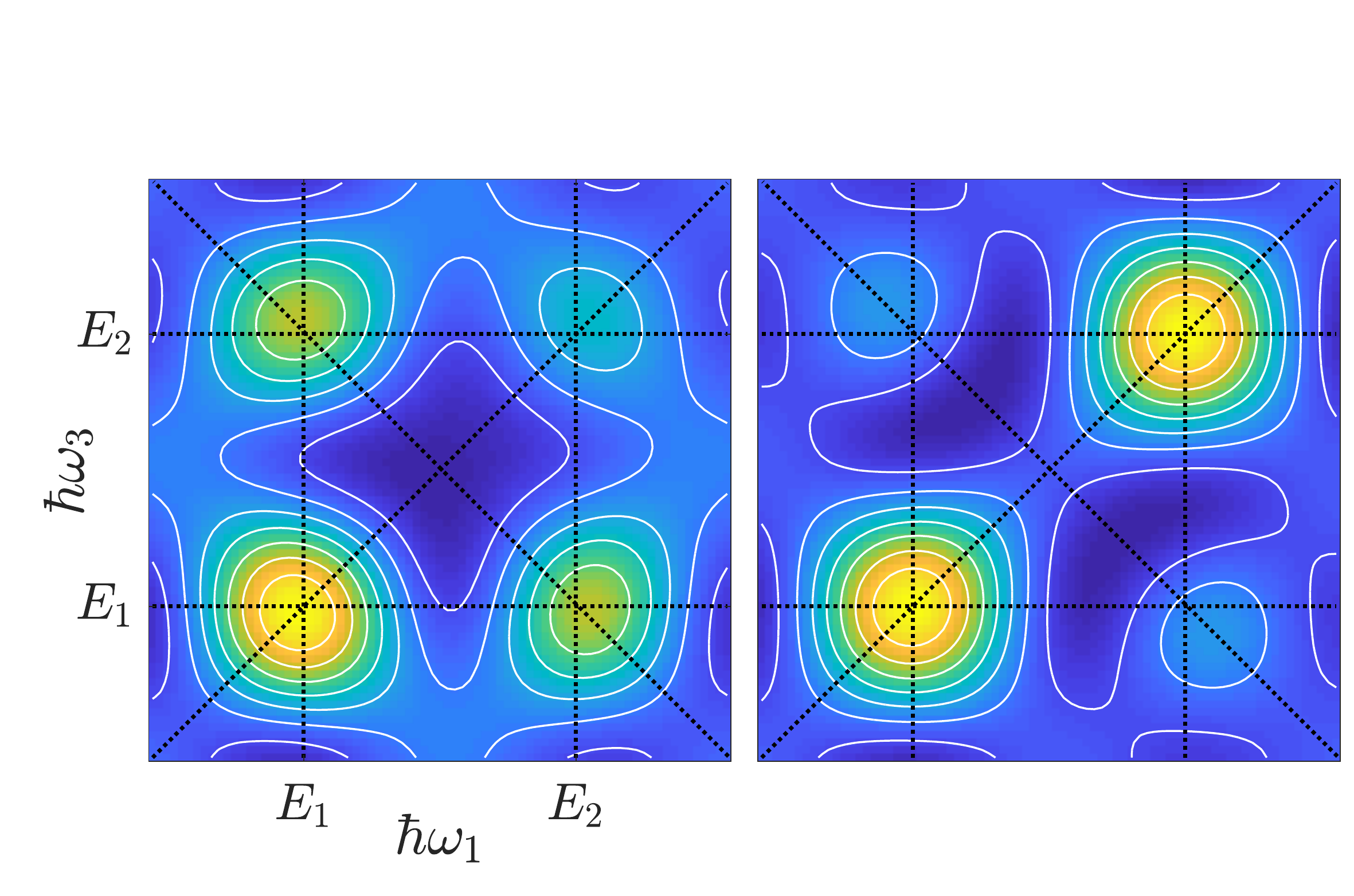}
         \put(-315,195){(a)}
         \put(-155,195){(b)}
\caption{\label{fig:fdhdDiagrams} The real parts of the zero-waiting time total 2D spectra, calculated with the traditional double-sided Feynman diagrams derived from perturbation theory. (a) the spectrum is simulated with pathways selected by the fluorescence detection setup, and (b) the heterodyne-detected counterpart is shown on the right. Note that the Fourier transform of the coherence time $t_1$ and signal time $t_3$ is taken for a time interval of 50 fs in order to achieve similar broadening as for the non-perturbative simulation, due to the finite Fourier transform.}
\end{figure}

For comparison, and as a starting point for the discussion of the 
effect of a non-idealised pulse, we compute the corresponding spectra using the double-sided Feynman diagrams laid out in ref. \citenum{PhysRevA.96.053830}, see figure~\ref{fig:fdhdDiagrams}. 


It is evident that the cross peaks in the HD spectrum are weaker than in the FD spectrum as would be expected by investigating the phase evolutions of the diagrams which reveal cancellations for the HD cross-peak diagrams. In fact, the cross peaks in the HD spectrum turn out to be artifacts of finite Fourier transforms of the first ($t_1=50$ fs) and last ($t_3=50$ fs) pulse delay, and disappear as these are increased. The reason for using finite Fourier transforms is simply to match the broadening arising from finite Fourier transforms in the non-perturbative simulations below.




Before we discuss the response to changes in the pulse amplitude, we note two general differences between FD and HD that we observe in figure~\ref{fig:fdhdDiagrams}: 1) The diagonal peaks are equally intense in the HD spectra, whereas the FD counterparts are biased to the lower diagonal peak. The explanation is that our model, as in reality, collects fluorescence emanating from the lowest excited state; this favours the lower diagonal peak over the upper diagonal which relies on the appropriate pathways to relax to the lowest excited state prior to fluorescing. In other words, the upper diagonal grows in as the fluorescence integration time is increased. For the HD version, both the highest and lowest excited states are detected, and the symmetry is only broken by the relaxation and dephasing processes. 2) The cross peak amplitudes are stronger for FD than for HD. This discrepancy is a direct consequence of the additional pathways selected by the FD scheme, specifically by excited state absorption (ESA) processes where the fourth pulse brings the state to a higher-lying population. The remaining pathways where the frequency evolutions correspond to a cross peak are equivalent for FD and HD and largely cancel out due to pairs of pathways with opposite phase progressions, but the cancellation is incomplete in non-idealised realisations, i.e. real-life experiments and simulations with finite pulses.

To calibrate our non-perturbative simulations we begin by finding appropriate pulse amplitudes for each detection scheme. For FD, we start off by replicating the calculation performed by Damtie et al\cite{PhysRevA.96.053830}., but with phase-cycling replacing phase-modulation. As can be seen in the appendix, we were able to produce the same spectra with our phase-cycling method using the same parameters, specifically a peak interaction energy of 8 meV, thus validating our approach and finding a suitable starting point for the pulse amplitude.



Intuitively, one might expect the HD and FD versions to operate within the same parameter space and to be similarly affected by changes to the experiment and system variables/parameters. However, when we repeated the simulation with the HD model, keeping all common parameters identical, we found that the two methods do not share the same parameter regime for the extraction of third-order signals. Specifically, the amplitude of the pulses required a sevenfold increase for easy acquisition of the HD signal, i.e.\ without an exceedingly high number of randomly positioned absorbers. Note that the HD third-order signal is a function of the number of absorbers (and their positions and orientations) so a large number of absorbers can to some degree compensate for low pulse amplitudes.


\subsection{Pulse Amplitude} 
\begin{figure}
         \includegraphics[width=\linewidth]{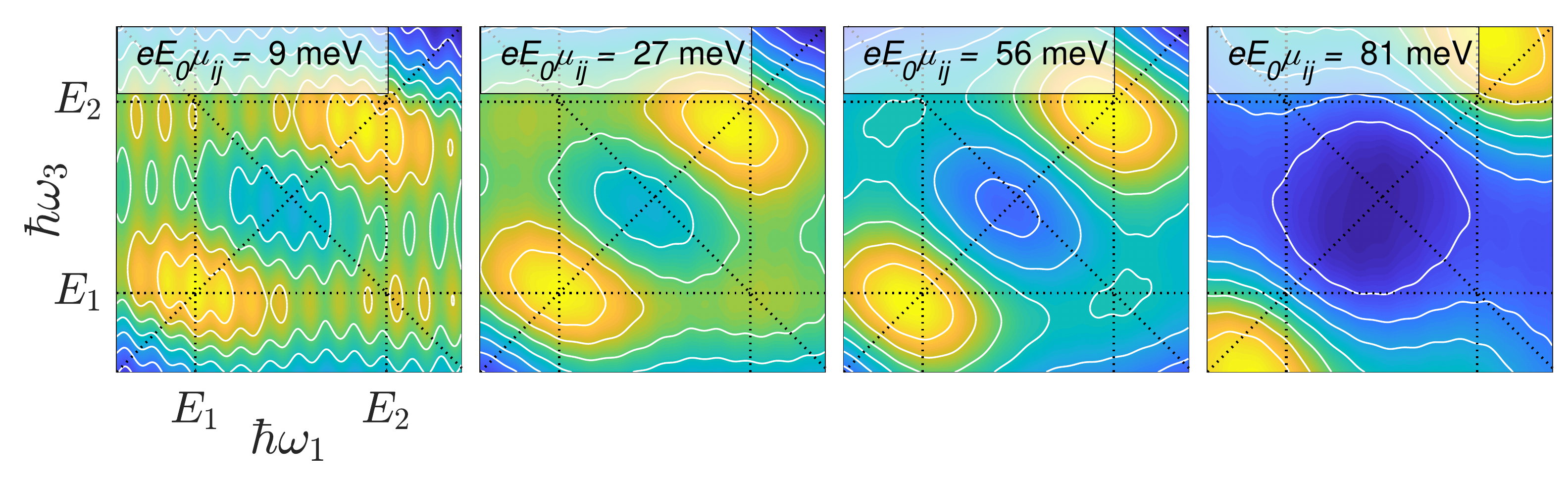}
\caption{\label{fig:hdAmplitude} The real parts of the zero-waiting time total 2D spectra using heterodyne-detection. The insets indicate the peak interaction energies, which were chosen to illustrate failure at low pulse amplitude (linear artifacts), at high pulse amplitude (higher-order effects) and the intermediate regime. The dotted lines guide the eye to the model system energies, the diagonal and antidiagonal. }
\end{figure}

One of the challenges that arises with explicit simulation of 2D spectra is that of the pulse/field power, which affects how much the state of the sample changes upon each interaction with a pulse. Unlike the pulse durations, which can be taken to be similar to the experimental standard or state-of-the-art, the pulse amplitudes must often be tweaked until a good signal is achieved.  If the power is too low or too high, it may be difficult to filter the desired third-order signal from the background of linear or higher orders. The window that is favourable for third-order detection may depend on how the signal is constructed. Given the different nature of the HD and FD schemes, and in particular how lower and higher orders are suppressed, this gives rise to a different range for the pulse amplitude. To illustrate this, we fix the pulse width by setting $\sigma=10$ fs and vary the pulse amplitude. The resulting total-correlation 2D spectra at zero waiting time for HD and FD are shown in figures \ref{fig:hdAmplitude} and \ref{fig:fdAmplitude}, respectively. 

It should be noted that the first and third pulse interval are scanned in steps of 10 fs up to 300 fs, which is much longer than the 50 fs interval used for Fourier transforming the corresponding intervals in the double-sided Feynman diagram calcaulations. This discrepancy stems from the fact that the undersampling, once folded within the Nyquist frequency appears to sample on a faster time scale. In our case the folding factor is 3, which means the undersampling at 10 fs corresponds to a normal sampling at $10-3\times \frac{2\pi \hbar}{\omega_{pulse}}= 1.73 $ fs. 30 steps of 1.73 fs adds up to about 50 fs. It follows that the decay from $|2\rangle$ to $|1\rangle$ will be greater with the longer scanning time. The reduced sensitivity by sampling beyond  1.3 times the decay rate\cite{Maciejewski2012} is counterbalanced by the increased frequency resolution and 300 fs is a good compromise.

\subsubsection{The effect on HD spectra} 

By sweeping the pulse amplitude across the third-order regime, see figure~\ref{fig:hdAmplitude}, we gain insight into the behaviour in the limits of linear and higher orders, as well as the intermediate regime. At the lowest pulse amplitude, linear artifacts appear in HD spectra as vertical streaks which translates as noise in the excitation frequency for a specific detection frequency. For an experienced experimenter or theoretician observing the spectra, such streaks would be met with suspicion and faced with scrutiny before they would be interpreted as a result of the underlying physics of the sample. In other words, these linear artifacts are not likely to be misinterpreted.

On the opposite end of the third-order regime, see the rightmost spectrum of figure~\ref{fig:hdAmplitude}, we see a perfectly plausible HD2D spectrum. It is only by comparison to the intermediate regime, the middle spectra of the same figure, that it is clear that the high pulse amplitude is introducing higher-order contributions into the spectrum. This demonstrates the importance of a well-calibrated pulse amplitude.

In the intermediate regime, see the middle spectra in figure \ref{fig:hdAmplitude}, we observe slight changes as the amplitude is increased. It can therefore be hard to justify quantitative conclusions based on a single 2D spectrum.  Particularly the cross peaks can be sensitive to the pulse amplitude. These arise from pathways with incorrect time-ordering, which are more likely at short waiting times. The zero waiting time therefore presents an additional challenge for the explicit simulation, as well as experiments carried out in the lab. 


\begin{figure}
\centering
         \includegraphics[width=0.45\linewidth]{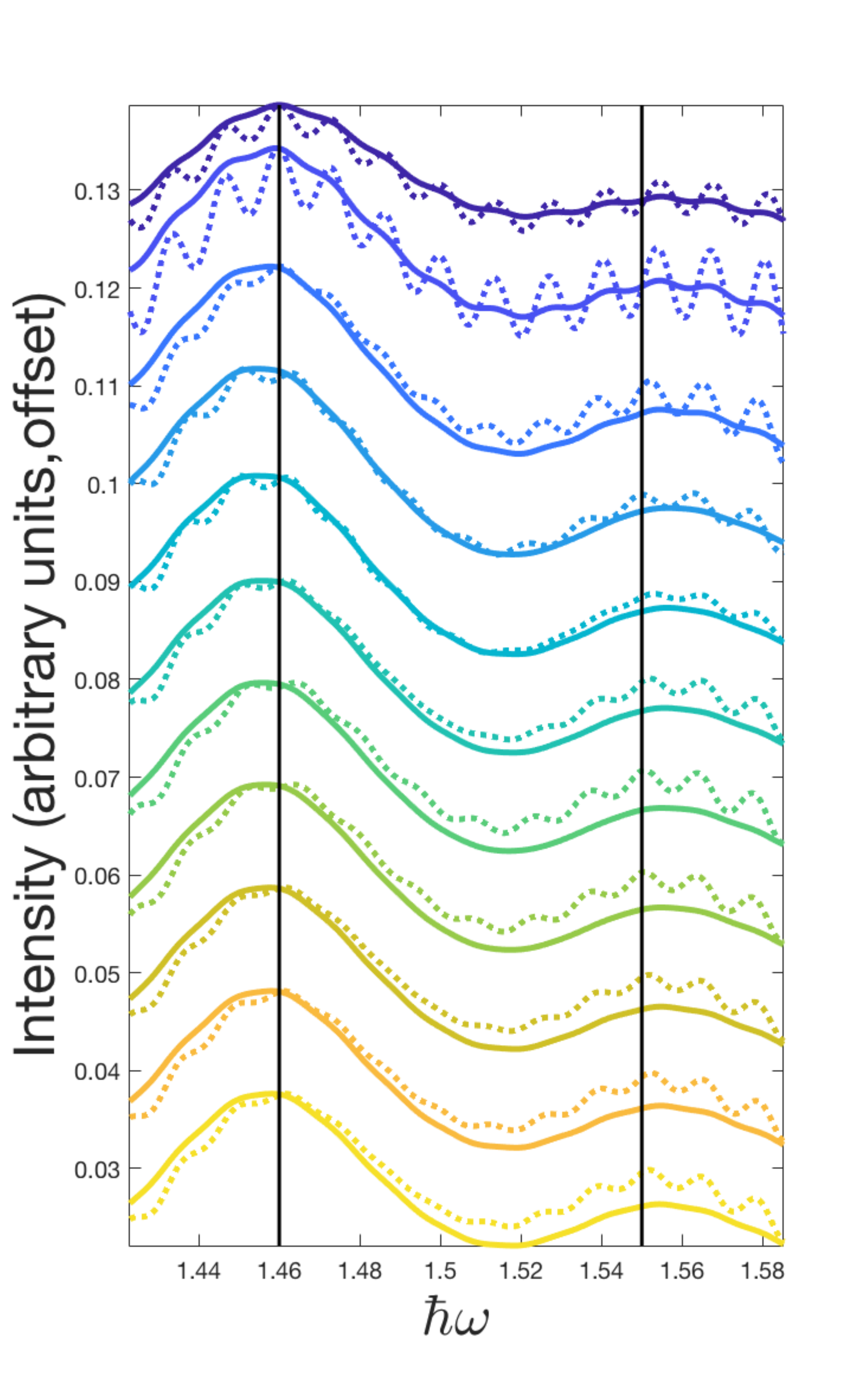}
\caption{\label{fig:fig4} The intensity along the horizontal line $\omega_3=E_1$ of a HD2D spectrum is plotted as a function of the waiting time in steps of 5 fs, with 0 fs waiting time at the top and 50 fs at the bottom. The full line is calculated using an intermediate pulse amplitude, corresponding to a peak interaction energy of 27 meV, and the dotted line is calculated using a low pulse amplitude with a peak interaction energy of 9 meV. The linear artifacts are strongest for the shortest waiting times such as 0 and 5 fs, with particularly the diagonal peak recovering as the waiting time is increased, but the effect is much less pronounced than for the FD case. Each pair of curves are normalised to the same maximum and are offset for visual clarity. }
\end{figure}

We also investigate the time evolution of the spectra as the waiting time is scanned in steps of 5 fs. Because the linear artifacts are clearly more pronounced in the horizontal axis, we pick the evolution of the $\omega_3 = E_1$ horizontal line to investigate whether the erratic spectrum at zero waiting time becomes smooth for longer waiting times. Figure \ref{fig:fig4} shows that there is a slight suppression of the linear artifacts as the waiting time is increased for the lower amplitude case, especially for the diagonal peak, but some noise still remains when compared to the intermediate amplitude regime.

\begin{figure}
         \includegraphics[width=\linewidth]{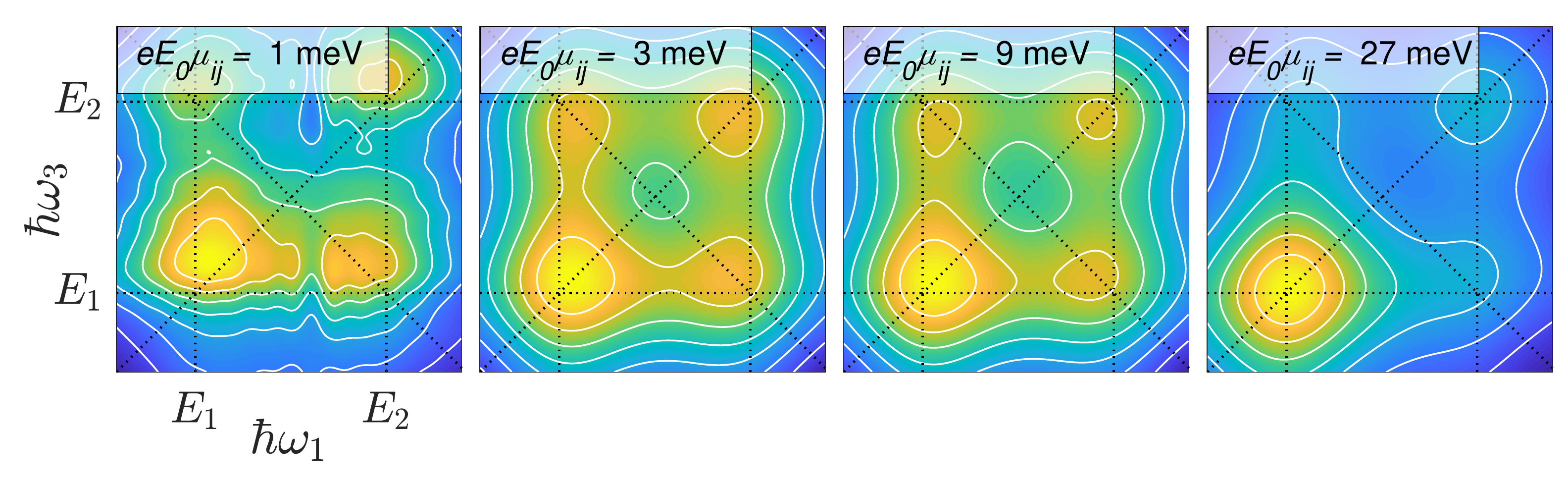}
\caption{\label{fig:fdAmplitude} The real parts of the zero-waiting time total 2D spectra using fluorescence-detection. The insets indicate the peak interaction energies, which were chosen to illustrate failure at low pulse amplitude (linear artifacts), at high pulse amplitude (higher-order effects) and the intermediate regime. The dotted lines guide the eye to the model system energies, the diagonal and antidiagonal. }
\end{figure}

\subsubsection{The effect on FD spectra} 

The effect of changes in the pulse amplitude on FD spectra is, similarly to the HD case, slight within the third-order regime, and strong at the limits of the regime. Interestingly, the way the spectra break down in these limits is in stark contrast to the HD case. At too low pulse amplitude, see the leftmost spectrum of figure~\ref{fig:fdAmplitude}, we do not observe linear artifacts as vertical streaks but instead the spectral peaks have a higher degree of randomness to them. At high pulse amplitudes, see the rightmost spectrum of figure~\ref{fig:fdAmplitude}, we see the lower diagonal peak gaining intensity and the upper diagonal peak losing intensity. The result is a spectrum which looks reasonable, but can lead to completely wrong analysis. 



\begin{figure}
\centering
         \includegraphics[width=0.45\linewidth]{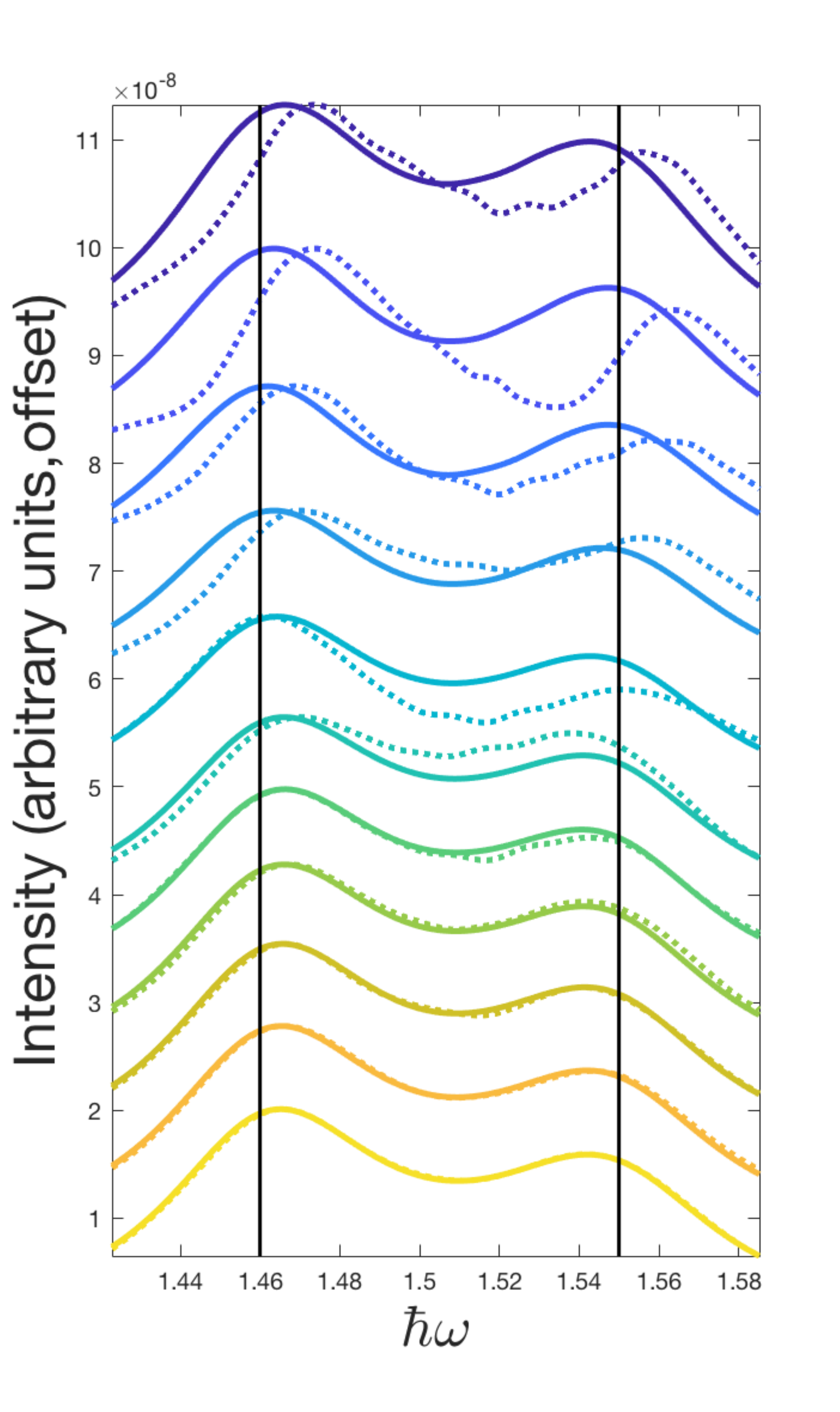}
\caption{\label{fig:fig3} The intensity on the diagonal line $\omega_1=\omega_3$ of a FD2D spectrum is plotted as a function of the waiting time in steps of 5 fs, with 0 fs waiting time at the top and 50 fs at the bottom. The full line is calculated using an intermediate pulse amplitude with a peak interaction energy of 3 meV, and the dotted line is calculated using a low pulse amplitude corresponding to a peak interaction energy of 1 meV. When the waiting time becomes large compared to the pulse width, the noise vanishes as a result of a diminishing contribution from incorrect time-ordering pathways. Each pair of curves are normalised to the same maximum and are offset for visual clarity.}
\end{figure}

We investigate the time evolution of the spectra by scanning the waiting time in steps of 5 fs. Figure \ref{fig:fig3} shows how the diagonal evolves for the low pulse amplitude case (corresponding to the leftmost spectrum in figure \ref{fig:fdAmplitude}) and for an intermediate pulse amplitude (corresponding to the second from the left in figure \ref{fig:fdAmplitude}). It is evident that, when the waiting time is increased, the low-amplitude case clears up and gradually resembles the intermediate amplitude case as the pulse overlap between the second and third pulse becomes negligible.



Comparing the two detection methods, we note that the HD third-order signal is more robust against higher-order contributions as the power is increased, as evidenced by the spectra acquired with pulse intensities corresponding to peak interaction energies of 27 and 56 meV, see the middle spectra in figure~\ref{fig:hdAmplitude}. The manner of which the third-order signal breaks apart in the limits of the third-order regime is very different, and stems from the disparate ways of how the third-order signal was constructed.


Note that amplitude affects the number of chromophores needed to phase the signal in HD2D. The comparison to FD2D is therefore not straightforward as only one chromophore is needed to compute its signal. HD tends to prefer a peak interaction energy upwards of 9 meV and can go to quite strong laser powers without compromising the third order signal, whereas FD is more prone to higher order effects, but still produces a clear signal at lower laser powers than HD.

\subsection{Pulse Durations} 

Seeing how influential the pulse overlap can be when the waiting time is varied, we proceed to discuss the effect of the pulse duration which will increase or decrease not only the overlaps between pulses 2 and 3 but potentially between all pulses.

For a fair comparison of the effect of the pulse duration, the pulse amplitude must be adjusted accordingly to ensure that the pulse power remains the same, thus keeping the population and coherence transfer rates on the same scale.

When the pulse duration is increased, the uncertainty in the excitation and detection frequencies also increases. To limit this effect, one can scan the respective pulse delays for longer. However, it is not possible to correct for the effects caused by pulse overlap. The more the pulses overlap, the stronger the signal from incorrect pulse ordering becomes, particularly in polarisation-controlled variants of 2DES as shown by Pale\v{c}ek et al.\cite{doi:10.1063/1.5079817} Care should therefore be taken when analysing experiments with considerable pulse overlap, especially of the early-time dynamics where pulse 2 and 3 are close together.

\begin{figure}
\centering
         \includegraphics[width=\linewidth]{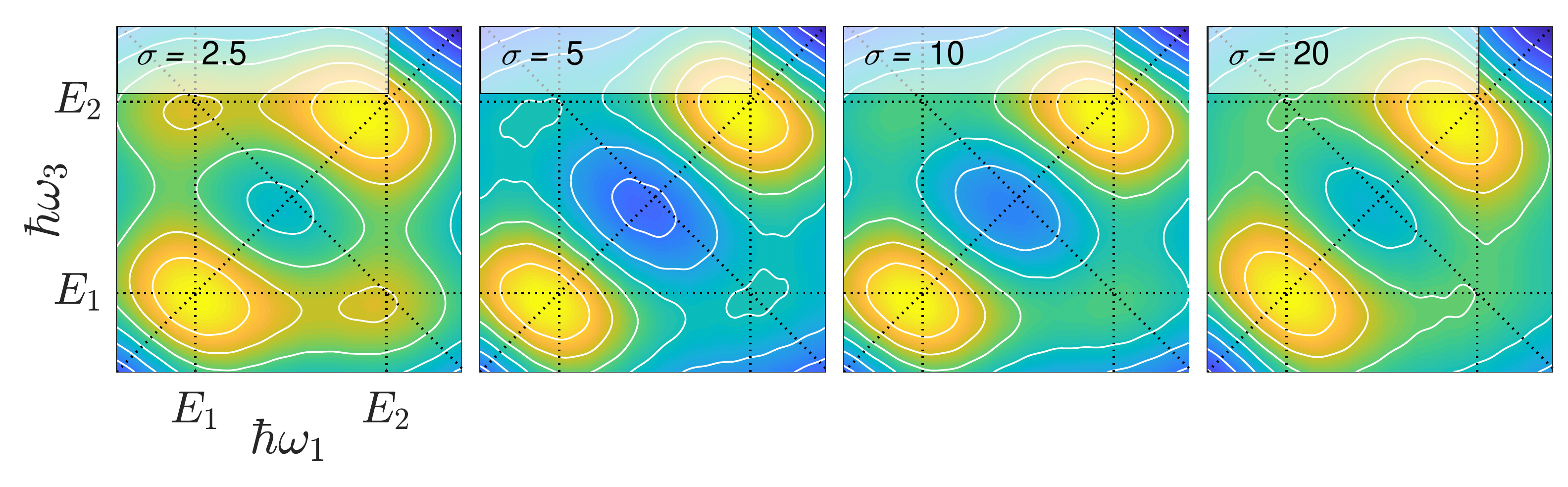}
\caption{\label{fig:hdDuration} The real parts of the zero-waiting time total 2D spectra using heterodyne-detection. The insets indicate the standard deviation of the Gaussian in fs, and hence the duration of the pulses. The $\sigma$ values range from 2.5 fs to 20 fs, which is spanning what is currently not possible to what is easily achieved in most 2DES setups. The dotted lines guide the eye to the model system energies, the diagonal and antidiagonal. }
\end{figure}

\begin{figure}
\centering
         \includegraphics[width=\linewidth]{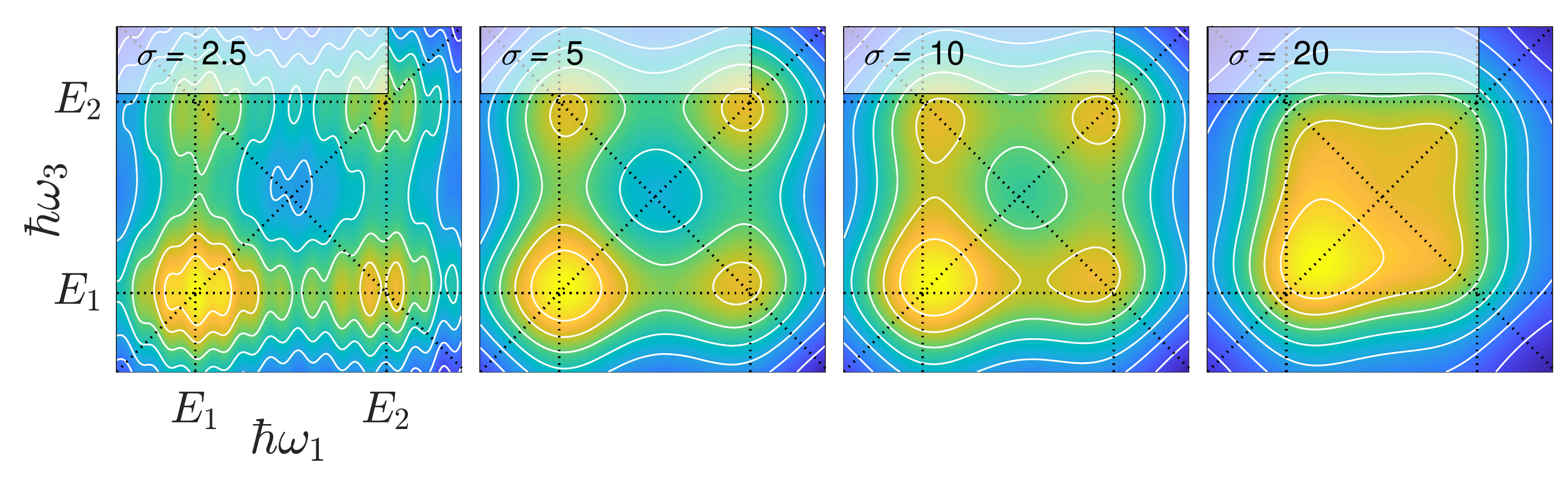}
\caption{\label{fig:fdDuration} The real parts of the zero-waiting time total 2D spectra using fluorescence-detection. The insets indicate the standard deviation of the Gaussian in fs, and hence the duration of the pulses. The $\sigma$ values range from 2.5 fs to 20 fs, which is spanning what is currently not possible to what is easily achieved in most 2DES setups. The dotted lines guide the eye to the model system energies, the diagonal and antidiagonal. }
\end{figure}

Figures~\ref{fig:hdDuration} and \ref{fig:fdDuration} show the total correlation 2D spectra at zero waiting time using FD and HD, respectively. Interestingly, HD is more robust against changes in the pulse duration. FD is more easily smeared and also struggles as the duration becomes very short, although this may be of little experimental interest as such pulse durations are not realistic/possible. From a theoretical standpoint, it is interesting to determine why the two detection methods differ. In HD, the phase picked up by each chromophore is solely by virtue of its position, whereas the phases picked up in FD are contained in the waveform and may be distorted as the pulse duration becomes comparable to the period of the pulse frequency. In any case, the linear artifacts seen for FD at the lowest pulse duration disappear when the waiting time is increased, just as we observed for HD at low pulse amplitude.

\section{Conclusions}

In our modelling of FD and HD 2DES we used pulses with variable amplitudes and durations. Within the semi-classical approximation, this approach can be considered a full model spectroscopy-wise. We observe effects related to the finite pulses, which can not be accounted for with the commonly employed Feynman diagram method. Of particular interest is the fact that the optimal window for the pulse power is different for the two detection schemes. The underlying reason for this discrepancy is the disparate ways that the third-order signal is constructed from the raw signal data. It is also interesting from a theoretical point of view to observe the dissimilar behaviour of the two methods as the limits of pulse amplitude and pulse duration are tested.

Computationally, the FD simulation is quite cheap because only 1 model system is required to create the 2D spectra, although 27 runs are required for phase-cycling. For HD, the number of simulation runs can be much higher, starting from hundreds but potentially requiring 10,000-100,000, depending on the ``phasing'' conditions (dephasing, timestep, scanning length etc.)

The investigation of the effect of the pulse amplitude and the pulse duration shows substantial and non-trivial changes within the third-order regime which calls for great care whenever quantitative analyses are attempted. Conclusions from quantitative analysis should essentially be backed up by non-perturbative simulations to take pulse effects into account. This is even more important when probing short-time dynamics where increased pulse overlapping is detrimental to the selection of spectroscopic pathways. 

On top of the complementary information found in FD2D spectra, much of the promise of FD2D is the fact that it can be contrasted against HD2D spectra to provide information that is not contained in either FD2D or HD2D. Therefore, it is imperative that all aspects of the experiment are well understood.

\section{Acknowledgements}
The authors acknowledge support of the Australian Research Council through grant CE170100026. This research was undertaken with the assistance of resources from the National Computational Infrastructure, which is supported by the Australian Government.

\appendix

\section{Comparison to Phase-Modulated FD Spectra}{}

\begin{figure}
\makebox[\textwidth][c]{\includegraphics[width=1.2\textwidth]{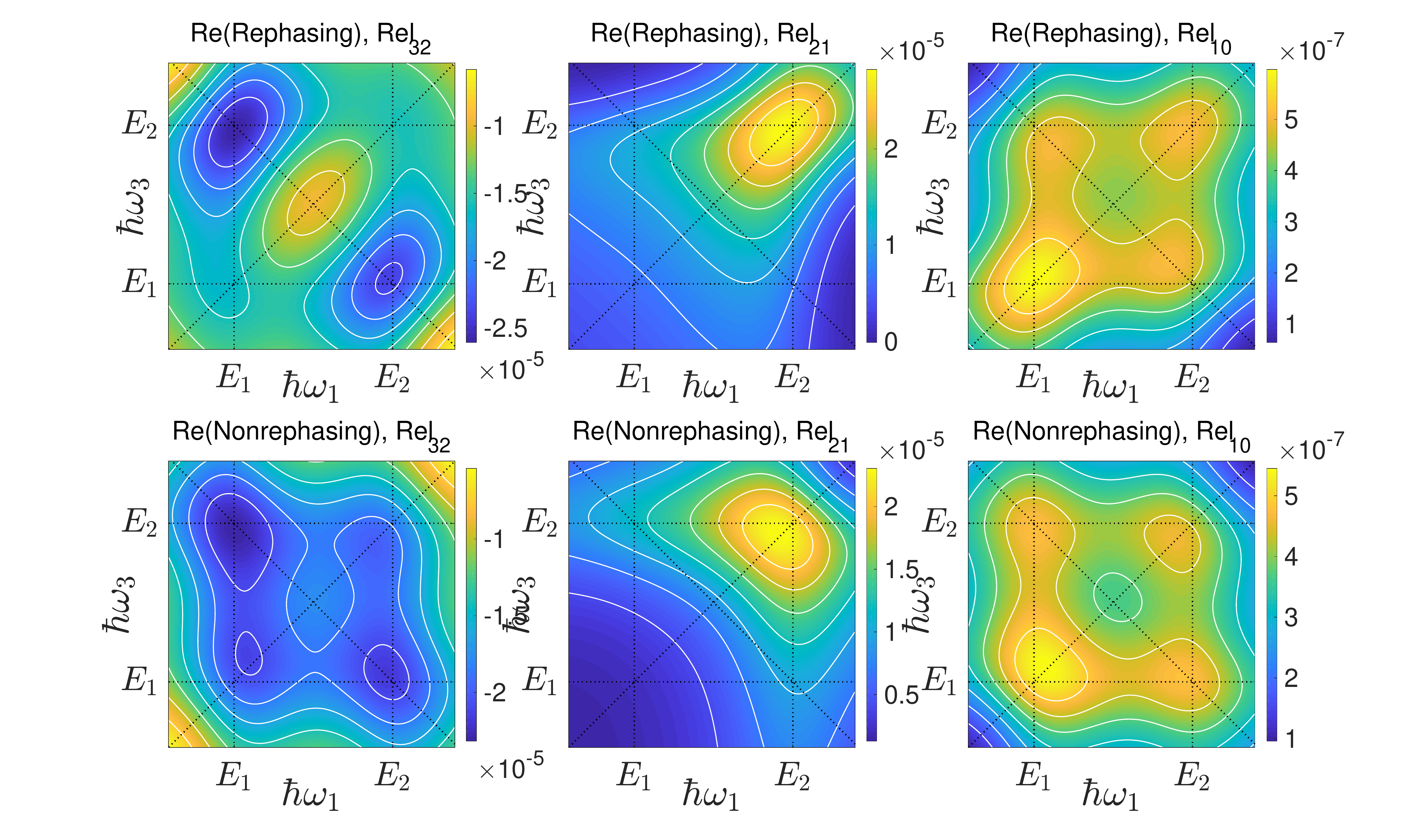}}%

\caption{\label{fig:replica} A replica of the zero-waiting time fluorescence-detected 2D spectra in ref. \citenum{PhysRevA.96.053830}. The only difference is that here phase-cycling is employed, instead of phase-modulation, as a method to filter the desired third-order signals from the background. Rel$_{ij}$ denotes the relaxation from $|i\rangle$ to $|j\rangle$, where Rel$_{10}$ is of primary relevance to fluorescence detection. Note that the sign is deliberately changed to promote the similarity to heterodyne-detected spectra, and that the axes are switched with respect to ref. \citenum{PhysRevA.96.053830} }
\end{figure}

By adopting the model from reference~\citenum{PhysRevA.96.053830}, we are able to verify that our calculations, using phase-cycling instead of phase-modulation, are correct. An alternative outcome could be that the two phasing schemes bring about differences in the spectra, but this turns out not to be the case as the resemblance is striking, see figure \ref{fig:replica} and figure 4 in the original work. Apart from replacing phase-modulation with phase-cycling, all parameters and equations are exactly the same.

For completeness, and as a starting point for the discussion on differences between explicitly simulated heterodyne-detected and fluorescence-detected 2D spectra, we also compute the HD spectra with the same parameters. However, in calculations of HD spectra, one must also decide on how many chromophores to include (and their positions) which will generally be iterated until convergence is achieved. This number can depend strongly on model parameters such as dephasing and energy levels, and on the specific realisations of the pulse trains, such as the pulse durations, pulse amplitudes, the scanning times and the number of scanning steps.

\begin{figure}
\makebox[\textwidth][c]{\includegraphics[width=1.0\textwidth]{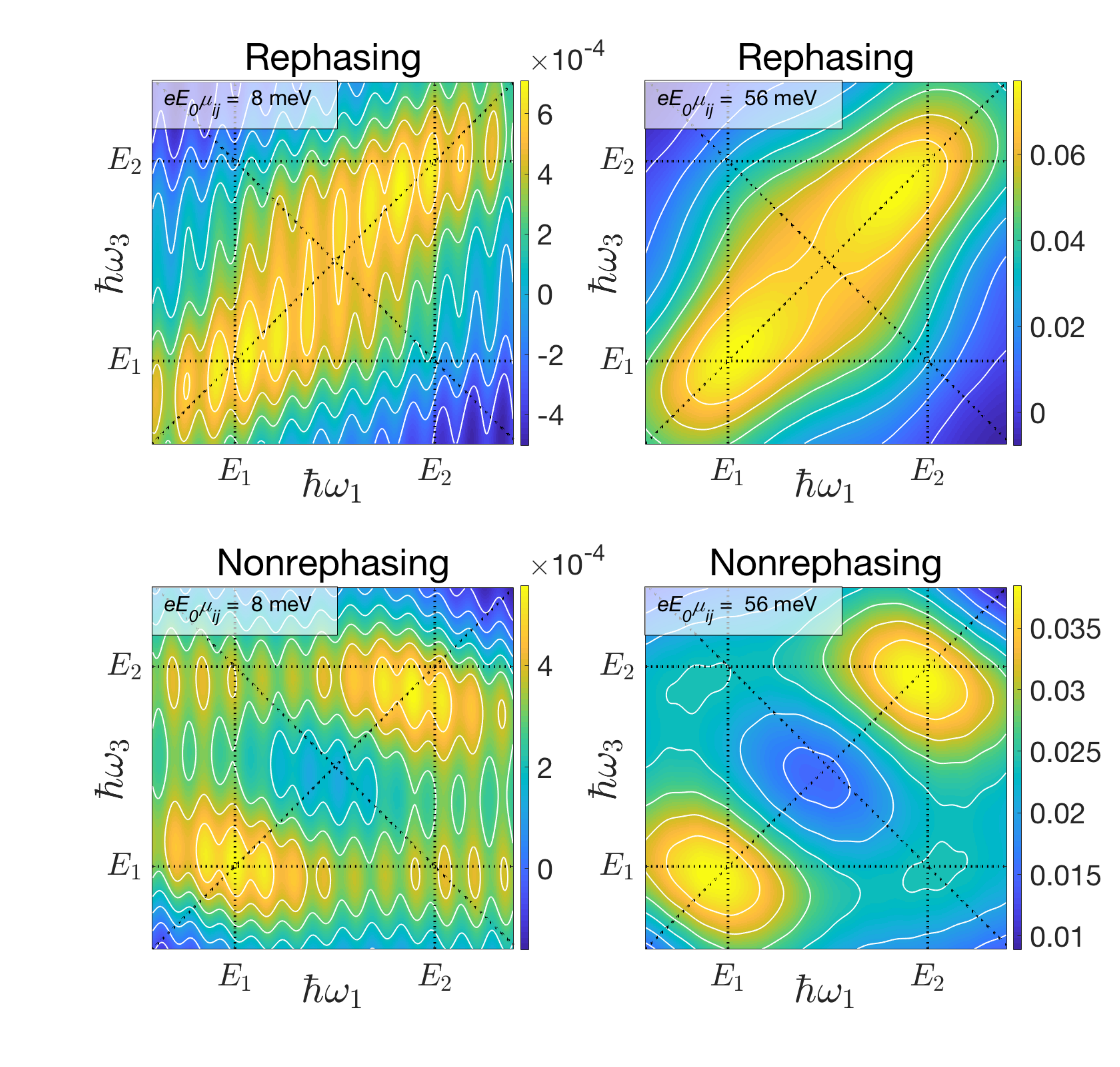}}%

\caption{\label{fig:hd_replica} Zero-waiting time heterodyne-detected 2D spectra. Left: The rephasing and nonrephasing signals were calculated using the same parameters as in ref. \citenum{PhysRevA.96.053830}. Right: The same parameters were used, apart from the pulse amplitude which was increased from 8 meV to 56 meV in peak interaction energy. 10,000 randomised positions were employed to converge the phasing of the third-order signal.}
\end{figure}

Intuitively, one might expect that the two detection schemes operate in the same field-strength range, meaning that the third-order signal starts to appear at the same laser-power threshold and that higher orders enter at similar intensities. However, the calculated spectra in figure \ref{fig:hd_replica} show that the pulse amplitude had to be increased sevenfold to achieve optimal third-order signals for the heterodyne-detected simulation. 

\newpage

\bibliographystyle{unsrt}
\bibliography{Paper2DHDFD}

\end{document}